\RequirePackage{lineno}

\documentclass[aps,pre,oldlfont,amsmath,amsfonts,amssymb,twocolumn]{revtex4}
\usepackage{bm}        

\usepackage[latin1]{inputenc}
\usepackage{graphicx}
\usepackage[english]{babel}
\usepackage[usenames,dvipsnames]{color}

\usepackage{amsfonts}
\usepackage{amsmath}
\usepackage{amssymb}

\usepackage{color}

\usepackage{soul}

\newcommand{\beq}{\begin{equation}}
\newcommand{\eeq}{\end{equation}}
\newcommand{\nn}{\nonumber}
\newcommand{\ket}[1]{|#1\rangle}
\newcommand{\bra}[1]{\langle #1|}

\newcommand{\ra}{\rightarrow}

\makeatletter
\@ifundefined{textcolor}{}
{%
 \definecolor{BLACK}{gray}{0}
 \definecolor{WHITE}{gray}{1}
 \definecolor{RED}{rgb}{1,0,0}
 \definecolor{GREEN}{rgb}{0,1,0}
 \definecolor{BLUE}{rgb}{0,0,1}
 \definecolor{CYAN}{cmyk}{1,0,0,0}
 \definecolor{MAGENTA}{cmyk}{0,1,0,0}
 \definecolor{YELLOW}{cmyk}{0,0,1,0}
 }

\begin{document}


\title{Quantum dissipative adaptation with cascaded photons}

\author{T. Ganascini
$^{1}$
}

\author{T. Werlang
$^{1}$
}

\author{D. Valente
$^{1}$
}
\email{valente.daniel@gmail.com}

\affiliation{
$^{1}$ 
Instituto de F\'isica, Universidade Federal de Mato Grosso, Cuiab\'a, MT, Brazil
}


\begin{abstract}
Classical dissipative adaptation is a hypothetical non-equilibrium thermodynamic principle of self-organization in driven matter, relating transition probabilities with the non-equilibrium work performed by an external drive on dissipative matter.
Recently, the dissipative adaptation hypothesis was extended to a quantum regime, with a theoretical model where only one single-photon pulse drives each atom of an ensemble.
Here, we further generalize that quantum model by analytically showing that N cascaded single-photon pulses driving each atom still fulfills a quantum dissipative adaptation.
Interestingly, we find that the level of self-organization achieved with two pulses can be matched with a single effective pulse only up to a threshold, above which the presence of more photons provide unparalleled degrees of self-organization.
\end{abstract}


\maketitle
\section{Introduction}

Finding nonequilibrium self-organization principles is a longstanding quest \cite{prigogine,anderson,JCP2013}.
The various behaviors of different nonequilibrium dissipative systems may seem too diverse to obey one set of principles.
Yet, as physicists we have witnessed many times in history where unified theories of seemingly distinct phenomena have come to light.
For instance, the Boltzmann distribution is a powerful principle relating the energy and the probability distribution for any system at thermal equilibrium with a weakly coupled reservoir.
Inspired by that, as well as by living matter \cite{JCP2013}, the principle of dissipative adaptation as a nonequilibrium version of the Boltzmann distribution has been recently proposed and discussed \cite{naturenano2015, PRL2017, ragazzon18, naturenano2020}.
According to the dissipative adaptation hypothesis, the transition probabilities between pairs of states are most relevant when a dissipative system is driven far from equilibrium, and the work performed by a given drive during the dynamical transition between the initial and the final state can dominate the probabilities with which these transitions occur.

\begin{figure}[!htb]
\centering
\includegraphics[width=1.0\linewidth]{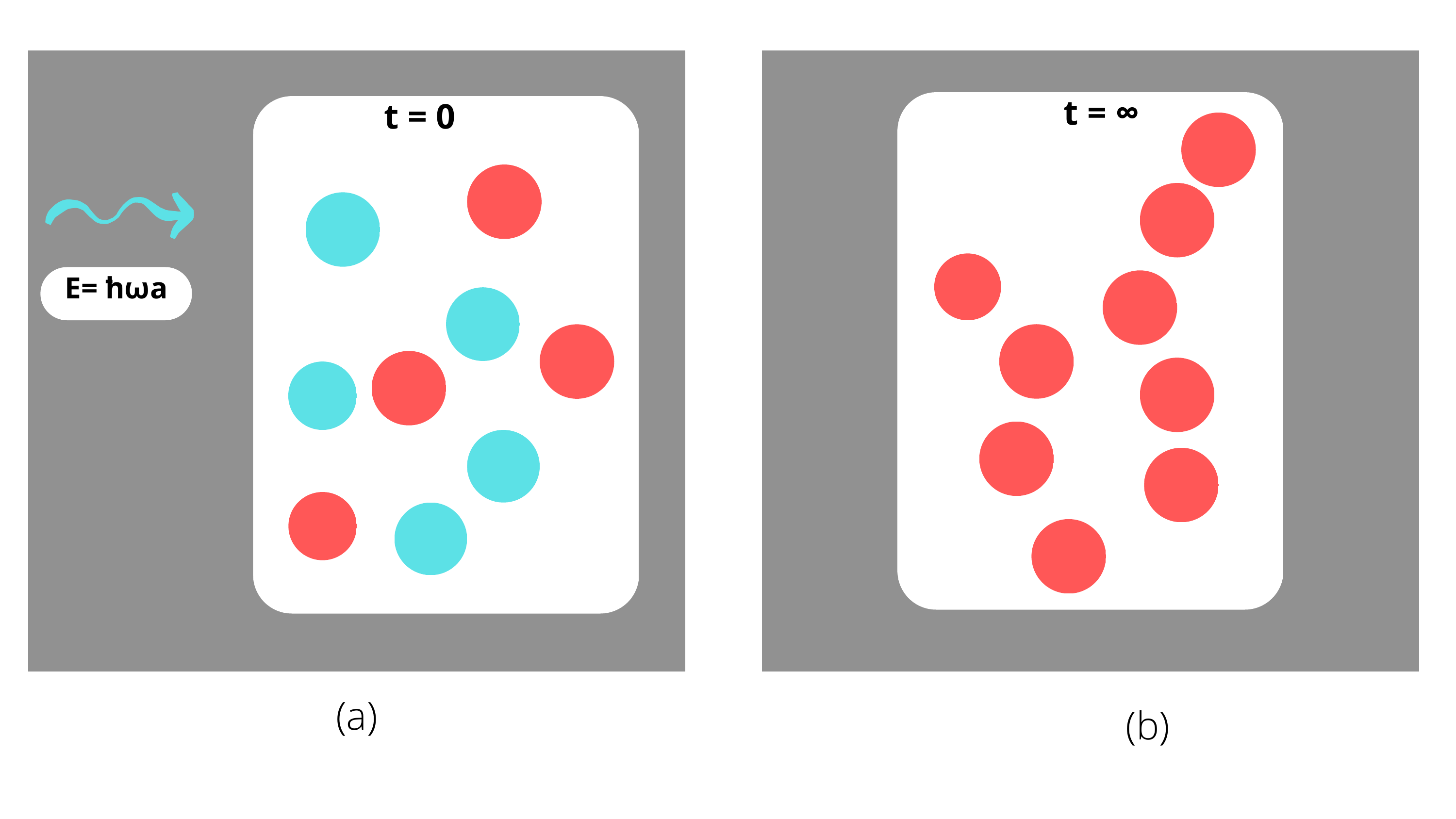} 
\caption{
The quantum self-organization as discussed here.
(a) Initially, at time $t=0$, an ensemble of $\Lambda$ atoms are in a mixed state, either at the ground state $\ket{a}$ (blue circles) or the ground state $\ket{b}$ (red circles), at temperature $T=0$.
Then, a nonequilibrium environment, composed of single-photon pulses (at resonance $\omega_L = \omega_a$), starts driving the atoms non-perturbatively.
(b) Eventually, after a long time $t\rightarrow \infty$, all the atoms transition to the self-organized (pure) state $\ket{b}$ (red circles).
Our goal in this paper is to understand the dynamics and thermodynamics of the process where $N$ photons drive each atom.
}
\label{fig1}
\end{figure}

To test whether the dissipative adaptation hypothesis is truly a general principle, we should look for it also at the most elementary scales of nature, where only few energy levels are present, temperatures are very low, and quantum coherences as well as quantum fluctuations are significant.
Indeed, a quantum dissipative adaptation has been recently found in the case where the source of work is a single-photon pulse, for which quantum fluctuations are far more significant than the average field \cite{qda,qsr}.
In its original formulation, the quantum dissipative adaptation also depends on quantum coherence.
A quite counterintuitive effect emerges in that case, namely, the highest average work performed by the photon does not correspond to the highest probability of single-photon absorption.
Finally, the quantum version is valid at zero temperatures, a regime forbidden by its classical counterpart.
Yet, both the classical and the quantum scenarios have shown evidence that work is a key thermodynamical quantity to bias the transition probabilities of a fluctuating system, hence determining the degree of stabilization (adaptation) to a given drive.
The sketch of the quantum self-organization described here can be seen in Fig.(\ref{fig1}).

Here, we raise the question of how broadly applicable is the idea of a quantum dissipative adaptation.
In particular, we are interested in the cases where quantum self-organization with a single photon is not ideal, i.e., when the transition probability between two ground states of a $\Lambda$ atom is less then unity after the single-photon pulse has gone away.
Is it true that the more energy is given (by means of more subsequent photon pulses arriving), the more the system adapts, so the higher the degree of self-organization?
That would provide a significant analogy with living matter, where we also see the increase of specialization in time provided enough resources.
This adds to a flourishing list of other analogies between living matter and dissipative systems, such as self-replication \cite{science2010,JCP2013,qsr} and the emergence of adaptive energy-seeking behavior \cite{kondepudi,cp22}; photonic systems included \cite{scirep13,natphot}.
In this paper, we show that the relation between the transition probability and the average work as found for a single photon remains valid if $N$ consecutive pulses of arbitrary shapes reach every atom in the ensemble.
Moreover, we show under which conditions a ``two-photon environment'' may improve the self-organization beyond any possible ``single-photon environment''.
Our results represent a further step in the direction of promoting the dissipative adaptation hypothesis to a general thermodynamic principle of non-equilibrium self-organization.

\section{Model}
We consider a global time-independent Hamiltonian of the system and its environment,
\beq
H=H_S + H_I + H_E.
\label{global}
\eeq
We model light-matter interaction via a dipole coupling, in the rotating-wave approximation \cite{qda, chen2004, dvnjp, dv2012, nonmarkov, OL2018},
\beq
H_I = -i\hbar \sum_\omega (g_a a_\omega \sigma_a^\dagger + g_b b_\omega \sigma_b^\dagger - \mbox{H.c.}),
\label{HI}
\eeq
where $\sigma_k = \ket{k}\bra{e}$ (for $k=a,b$) and $\mbox{H.c.}$ is the Hermitian conjugate.
Orthogonal polarization modes $\left\{ a_\omega \right\} $ and $\left\{ b_\omega \right\}$ respectively interact with transitions $\ket{a}$ to $\ket{e}$ and $\ket{b}$ to $\ket{e}$.
This is a multi-mode Jaynes-Cummings model, and our interest is in the limit of a broadband continuum of frequencies, 
\beq
\sum_\omega \ra \int d\omega \varrho_\omega \approx \varrho \int d\omega,
\eeq
where we have defined a constant density of modes $\varrho_\omega \approx \varrho$.
Physically, this means that there is no frequency filter in our idealized environment.
The continuum of frequencies allows us to employ a Wigner-Weisskopf approximation to obtain dissipation rates
$\Gamma_{a} = 2\pi g_a^2 \varrho$ and 
$\Gamma_{b} = 2\pi g_b^2 \varrho$.
The environment Hamiltonian reads
\beq
H_E = \sum_\omega \hbar \omega (a_\omega^\dagger a_\omega + b_\omega^\dagger b_\omega).
\eeq
The system is described by a three-level Hamiltonian, with states named $\ket{a}$, $\ket{b}$, and $\ket{e}$, so that
\beq
H_S = \hbar \omega_a \ket{e}\bra{e} + \hbar \delta_{ab} \ket{b}\bra{b}.
\eeq
Most importantly, when in $\Lambda$ configuration, these energy levels provide an elementary energy barrier (the excited state $\ket{e}$) separating two highly stable states (the two ground states $\ket{a}$ and $\ket{b}$).
The energy difference between states $\ket{b}$ and $\ket{a}$, given by $\hbar\delta_{ab} = \hbar(\omega_a - \omega_b)$, although relevant for thermal equilibrium considerations, does not affect our results as far as nonequilibrium effects are concerned.

\section{Results}
\subsection{Revisiting the single-photon wavepacket scenario}
In Ref.\cite{qda}, the initial state of the field has been considered as a single-photon pulse added to a zero-temperature environment, namely,
\beq
\ket{1_a} = \sum_\omega \phi_\omega^{a}(0) \ a_\omega^\dagger \ket{0},
\label{1a}
\eeq
where
$\ket{0} = \prod_\omega \ket{0_\omega^{a}} \otimes \ket{0_\omega^{b}}$ is the vacuum state of all the field modes.
Note that the polarization is fixed as that of the $a_\omega$ modes, thus imposing an environmental condition to which the atom is expected to adapt.
The level of adaptation will depend on the pulse shape, $\phi_\omega^{a}(0)$, which admits a spatial dependent representation,
\beq
\phi_a(z,t) = \sum_\omega \phi_\omega^{a}(t) e^{i k_\omega z}.
\eeq
For simplicity, our electromagnetic environment is one-dimensional in space, explaining why only coordinate $z$ appears.
Also, the modes propagate towards the positive direction, with a dispersion relation in the form $k_\omega = \omega /c$.
This idealized model closely approximates the experimental scenario known as waveguide quantum electrodynamics (waveguide-QED) \cite{lodahl, nori}.
If the photon is prepared by means of the spontaneous emission of a distant source atom (not shown in the model), then we would have that
\beq
\phi_a(z,0) = N \Theta(-z) \exp[(\Delta / 2 + i \omega_L)z/c],
\label{expa}
\eeq
where $\Delta$ is the linewidth of the pulse (the lifetime of the atomic source), and $\omega_L$ is the central frequency of the photon (the transition frequency of the atomic source).
The size of the pulse in space is characterized by $c\Delta^{-1}$, in this case.

The key quantities analyzed in the context of dissipative adaptation are transition probabilities and their relations with the absorbed work from an external drive.
In the quantum case, we are especially interested in the transition probability from $\ket{a}$ to $\ket{b}$,
\beq
p_{a\ra b}(t) \equiv \bra{b} \mbox{tr}_E \left[ U\ket{a, 1_a}\bra{a, 1_a}U^\dagger \right] \ket{b},
\label{pnk}
\eeq
where $U = \exp(-iHt/\hbar)$, and $H$ is the global Hamiltonian in Eq.(\ref{global}), and $\mbox{tr}_E$ is the partial trace over the field modes.
Taking advantage of the number conservation in the rotating-wave approximation, we restrict our model to the single-excitation subspace, parametrized by
$U\ket{a,1_a} = \psi(t) \ket{e,0} 
+ \sum_\omega \phi_\omega^a(t) a_\omega^\dagger \ket{a,0}
+ \phi_\omega^b(t) b_\omega^\dagger \ket{b,0}$.
We now have that
\begin{eqnarray}
p_{a\ra b}(t) &=& \sum_\omega |\phi_\omega^b(t)|^2 \nn\\
&=& (2\pi \varrho c)^{-1} \int_{-\infty}^{\infty} dz |\phi_b(z,t)|^2.
\label{pnk2}
\end{eqnarray}

To obtain the probability amplitude of the field, we make a Wigner-Weisskopf approximation, thus obtaining that
\begin{align}
\phi_b(z,t) &= \phi_b(z-ct,0) e^{-i\delta_{ab} t} \nn \\
&+ \sqrt{2\pi \varrho \Gamma_b} \Theta(z) \Theta(t-z/c) \psi(t-z/c) e^{-i\delta_{ab} z/c},
\label{phib}
\end{align}
where $\Theta(z)$ is the Heaviside step function.
This expression evidences the superposition of the freely propagating driving field with that emitted by the atom.
Because the incoming pulse is fixed to the $a_\omega$ modes, we have that $\phi_b(z-ct,0) = 0$.
By using Eqs.(\ref{phib}) and (\ref{pnk2}), we find that
\beq
p_{a\ra b}(t) = \Gamma_b \int_0^t dt' |\psi(t')|^2,
\label{history}
\eeq
after an appropriate change of variables.
The meaning of Eq.(\ref{history}) is that, for the system to get to state $\ket{b}$ departing from $\ket{a}$, the entire history of the excitation amplitude from time $0$ to $t$ matters.

We can think of two options for maximizing that integration: 
(1) by increasing the excitation of the atom (making $|\psi(t)|$ as large as possible); or 
(2) by increasing the time duration of the pulse (without necessarily exciting the atom too much, $ |\psi(t)| \ll 1 $).

To test options (1) and (2) above, we have to solve for the excitation amplitude.
Employing once again the Wigner-Weisskopf approximation, we have that
\beq
\partial_t \psi(t) =  - \left( \frac{\Gamma_a + \Gamma_b}{2} + i\omega_a \right) \psi(t) - g_a \phi_a(-ct,0).
\label{eqpsi}
\eeq
For $\psi(0)=0$, this gives us
\beq
\psi(t) = - g_a \int_0^t \phi_a(-ct',0) e^{-\left( \frac{\Gamma_a+\Gamma_b}{2} + i\omega_a \right)(t-t')} dt',
\label{psi}
\eeq
which provides a generic solution.

For the sake of definiteness, let us take the exponential profile from Eq.(\ref{expa}), which gives
\beq
\psi(t) = 
- A 
e^{-\left( \frac{\Gamma_a + \Gamma_b}{2} + i \omega_a \right) t} 
\left[ e^{\left( \frac{\Gamma_a+\Gamma_b-\Delta}{2} - i\delta_L \right)t} - 1 \right],
\label{solpsi}
\eeq
where
\beq
A \equiv \frac{\sqrt{\Gamma_a \Delta}}{\frac{\Gamma_a+\Gamma_b - \Delta}{2} - i \delta_L},
\eeq
with $\delta_L \equiv \omega_L - \omega_a$.
The resonance condition, $\delta_L = 0$, is necessary to maximize $|\psi(t)|$.

Now the relevant degree of freedom is the size of the pulse, $\Delta^{-1}$, given fixed $\Gamma_a$ and $\Gamma_b$.
If $\Delta^{-1} \rightarrow \infty$, the pulse is very long and the excitation is arbitrarily small, $|\psi(t)|^2 \propto \Delta \rightarrow 0$, as discussed in option (2).
If $\Delta^{-1} \rightarrow 0$, the pulse is arbitrarily short, also making the excitation probability to vanish, 
$|\psi(t)|^2 \propto \Delta^{-1} \rightarrow 0$.
This is the worst case scenario, since both the size and the duration of the atomic excitation are negligible.
For intermediate pulse sizes, $\Delta \sim \Gamma_a \sim \Gamma_b$, we find from Eq.(\ref{solpsi}) that
\beq
|\psi(t)| \leq 2 \sqrt{ \frac{\Gamma_a}{\Gamma} } \left( \frac{\Delta}{\Gamma} \right)^{- \frac{1}{2}\frac{\Delta+\Gamma}{\Delta - \Gamma} },
\eeq
where we have defined $\Gamma = \Gamma_a + \Gamma_b$.
The excitation probability is therefore maximized at $\Delta = \Gamma$, when the duration of the pulse is equal to the atomic lifetime.
This is the scenario of option (1).

By substituting Eq.(\ref{solpsi}) into (\ref{history}), we find that
\beq
p_{a\ra b}(\infty) = \frac{4\Gamma_a \Gamma_b}{\Gamma(\Gamma+\Delta)}.
\label{pabDelta}
\eeq
This clearly shows that option (2) is the correct one.
That is, the transition probability is maximized when the duration of the pulse is maximal ($\Delta/\Gamma \rightarrow 0$), even though the excitation probability is vanishingly small in that same situation.
This is a signature of the beneficial effect of quantum coherence in this process.
However, this raises the question: if the excitation is negligible, is the work transferred from the photon to the atom also negligible, therefore violating the classical dissipative adaptation hypothesis?

To answer that question, we have to define the work performed by the single-photon drive.
Classically, the work of a time-varying classical electric field $E(t)$ acting on a classical dipole $D(t)=qx(t)$, is given by 
$W  = \int F \dot{x} \ dt = \int q E \dot{x} \ dt = \int \dot{D} E \ dt$.
Here, we define the average work performed by a single-photon pulse on a quantum dipole by employing the Heisenberg picture,
\begin{equation}
\langle W \rangle = \int_0^\infty \langle \left( \partial_t D(t) \right) E_{\mathrm{in}}(t) \rangle \ dt,
\label{defw}
\end{equation}
where $E_{\mathrm{in}}(t) = \sum_\omega i \epsilon_\omega a_\omega e^{-i\omega t} + \mbox{h.c.}$ is the incoming field.
The field produced by the atom that acts back on the atom itself gives rise to heat dissipation in our model.
The dipole operator is given by $D(t) = U^\dagger D U$, with 
$D = \sum D_{ea} \sigma_a + \mbox{h.c.}$, so that $\hbar g_a = D_{ea} \epsilon_{\omega_a}$.

Using integration by parts, we can rewrite Eq.(\ref{defw}) as 
$\langle W \rangle = -\int_0^\infty  \langle D(t) \partial_t E_{\mathrm{in}}(t) \rangle dt$.
Within the rotating-wave approximation, this gives us that
\begin{align}
\langle W\rangle = -\int_0^\infty dt \ &(i\hbar) \sum_\omega (-i\omega) g_a \langle \sigma_{a}^\dagger(t) a_\omega \rangle e^{-i\omega t} \nn\\
&+ (-i\omega) g_b \langle \sigma_{b}^\dagger(t) b_\omega \rangle e^{-i\omega t} \nn\\
&+ \mbox{c.c.},
\end{align}
where $\mbox{c.c}$ stands for complex conjugate.
Choosing $\ket{1_a}$ as the initial state of the field implies that $\langle \sigma_{b}^\dagger(t) b_\omega \rangle = 0$.
The non-zero correlation function is
\begin{align}
\langle \sigma_{a}^\dagger(t) a_\omega \rangle 
&= \langle U^\dagger \sigma_{a}^\dagger U  a_\omega \ket{a, 1_a} \\
&= \langle U^\dagger \sigma_{a}^\dagger U  \phi_\omega(0) \ket{a,0} \\
&= \phi_\omega(0) \bra{a,1_a} U^\dagger \ket{e,0} \\
&= \phi_\omega(0) \psi^*(t).
\end{align}
We thus get that,
\beq
\langle W \rangle = 
-\hbar g_a \int_0^\infty dt \psi^*(t)  i \partial_t \phi_a(-ct,0)  + \mbox{c.c.}.
\label{windetail}
\eeq
For a pulse of central frequency $\omega_L$ and general envelope shape, we define $\phi_a(-ct,0) = \phi_a^{env}(-ct,0)\exp(-i\omega_L t)$.
The derivative of the fast oscillating part gives rise to
\beq
\langle W \rangle = 
\hbar \omega_L \int_0^\infty dt \ (-2g_a \mbox{Re}[\psi^*(t) \phi_a(-ct,0)]).
\label{w1}
\eeq
The derivative of the slowly varying part is related with $\mbox{Im}[\psi^*(t) \exp(-i\omega_L t)]$, since $\partial_t \phi_a^{env}(t)$ is real.
From Eq.(\ref{psi}), it follows that $\mbox{Im}[\psi^*(t) \exp(-i\omega_L t)] = 0$ if $\delta_L = 0$.
This shows that, at resonance, only the absorptive contribution remains, while the dispersive (reactive) contribution vanishes.

From Eq.(\ref{eqpsi}), we can derive the dynamics of the excited-state population $|\psi(t)|^2$,
\beq
\partial_t |\psi(t)|^2 = -\Gamma |\psi(t)|^2 - 2g_a \mbox{Re}[\psi^*(t) \phi_a(-ct,0)].
\eeq
Substituting it back in Eq.(\ref{w1}), gives us that
\beq
\langle W \rangle = 
\hbar \omega_L  
\int_0^\infty \Gamma |\psi(t)|^2 dt,
\label{qda}
\eeq
where we have used that $|\psi(0)|^2 = |\psi(\infty)|^2 = 0$.
By comparing Eqs.(\ref{qda}) and (\ref{history}), we immediately see that
\beq
p_{a\rightarrow b}(\infty) = \frac{\Gamma_b}{\Gamma} \ \frac{\langle W \rangle}{\hbar \omega_L},
\eeq
which is the quantum dissipative adaptation relation for a generic single-photon pulse (not restricted to the exponential profile).
This result answers to the question we have raised above by showing that, even though the excitation probability is negligible in the highly monochromatic case, the transferred work is maximized in that regime along with the transition probability from $\ket{a}$ to $\ket{b}$.

\subsection{Quantum dissipative adaptation for cascaded photons}
By cascaded interactions, we mean that consecutive non-overlapping and uncorrelated single-photon pulses are driving the atom.
Under this assumption, we have that, either the first photon promotes the transition, or it leaves the atom unaltered and the second photon enables the transition, and so on.
Asymptotically ($t\rightarrow \infty$), the total probability transition $p^{(T,N)}_{a\rightarrow b}$ after $N$ pulses have driven the atom is given by 
\beq
p^{(T,N)}_{a\rightarrow b} = \sum_{k=1}^{N} \prod_{j=1}^{k-1} p^{(j)}_{a\rightarrow a} p^{(k)}_{a\rightarrow b}.
\label{ptot}
\eeq
Here, $p^{(j)}_{a\rightarrow a} = 1-p^{(j)}_{a \rightarrow b}$ is the probability that the atom remains in state $\ket{a}$ during the interaction with the $j$-th pulse, and $p^{(k)}_{a\rightarrow b}$ is the transition probability due to the $k$-th photon.
The degrees of freedom considered in this model are the linewidths $\left\{\Delta_k \right\}$ of the $N$ photon pulses.

The average work in this cascaded process is additive,
\beq
\langle W^{(N)} \rangle = \sum_{k=1}^{N} \langle W \rangle_k,
\eeq
where $\langle W \rangle_k$ is the average work performed by the $k$-th photon of linewidth $\Delta_k$.
The average work performed by the $k$-th photon is given by
\beq
 \langle W \rangle_k = p^{(k-1)}_a  \langle W \rangle_a^{(k)} + p^{(k-1)}_b  \langle W \rangle_b^{(k)},
\eeq
which depends on the probability $p^{(k-1)}_i$ that the previous photon has left the atom in state $\ket{i}$.
Also, we have defined 
$\langle W \rangle_{a}^{(k)}$ 
to evidence that the initial state of the system for which the work is being computed is 
$\ket{a,1_a}$, and 
$\ket{b,1_a}$ for 
$\langle W \rangle_{b}^{(k)}$.
That is, all the photons are initially equally polarized, only the atom may be in a different state.
We thus find that $\langle W \rangle_{b}^{(k)} = 0$ for all $k$, implying that
\beq
\langle W^{(N)} \rangle = \sum_{k=1}^{N} p^{(k-1)}_a  \langle W \rangle_a^{(k)}. 
\eeq
Because there is no physical mechanism in our model that makes the atom jump back from $\ket{b}$ to $\ket{a}$ (due to the zero temperature), the only possible path for it to be found at $\ket{a}$ is that where the atom has never left state $\ket{a}$ throughout its entire history.
This means that $p^{(k-1)}_a$ is a product of the probabilities that all the previous photons have also left the atom at $\ket{a}$, so that
\beq
\langle W^{(N)} \rangle = \sum_{k=1}^{N} \left( \prod_{j=1}^{k-1} p^{(j)}_{a\rightarrow a} \right) 
 \langle W \rangle_a^{(k)}.
\eeq
Using the quantum dissipative adaptation relation for each single photon, namely, 
$\langle W \rangle_a^{(k)} = p_{a\rightarrow b}^{(k)} \hbar \omega_L \Gamma/\Gamma_b$,
we have that
\beq
\langle W^{(N)} \rangle 
= 
\left(
\sum_{k=1}^{N} \prod_{j=1}^{k-1} p^{(j)}_{a\rightarrow a} p_{a\rightarrow b}^{(k)} 
\right) \hbar \omega_L \Gamma/\Gamma_b.
\eeq
By employing Eq.(\ref{ptot}), we finally find that
\beq
p^{(T,N)}_{a\rightarrow b} = \frac{\Gamma_b}{\Gamma} \frac{\langle W^{(N)} \rangle}{\hbar \omega_L},
\eeq
showing that the quantum dissipative adaptation remains valid for $N$ cascaded photons of arbitrary shapes.

\subsection{Long versus short pulses}
Is it true that two pulses always increase the degree of organization of the atom? 
Or can we find a single photon long enough to produce an equivalent result?
With these questions in mind, we take a closer look at the $N=2$ case.

To answer that question, we analyze 
\begin{align}
p^{(T,2)}_{a\rightarrow b} 
&= p^{(1)}_{a\rightarrow a} p^{(2)}_{a\rightarrow b} + p^{(1)}_{a\rightarrow b},\nn \\
&= (1- p^{(1)}_{a\rightarrow b}) p^{(2)}_{a\rightarrow b} + p^{(1)}_{a\rightarrow b}.
\end{align}
We assume from here on that our photons have exponential envelope profiles, as described by Eq.(\ref{expa}), with two generally distinct linewidths $\Delta_1$ and $\Delta_2$.
Thus Eq.(\ref{pabDelta}) is valid, and we have that
\beq
p^{(k)}_{a\rightarrow b} = \frac{r}{1+\Delta_k/\Gamma},
\eeq
where $r = 4\Gamma_a\Gamma_b/\Gamma^2$, and $\Gamma=\Gamma_a + \Gamma_b$.

Let us take the limit of two pulses that are both very short in time and space (i.e., highly broadband), $\Delta_1 \gg \Gamma$ and $\Delta_2 \gg \Gamma$.
In that case,
\beq
p^{(k)}_{a\rightarrow b} \approx r\Gamma \tau_k \ll 1,
\eeq
where $\tau_k = \Delta_k^{-1}$ is the typical time duration of the pulse.
The total probability now reads
\beq
p^{(T,2)}_{a\rightarrow b} 
\approx r\Gamma (\tau_1 + \tau_2) 
= r\Gamma \tau_{\mathrm{eff}}
\approx p^{(1)}_{a\rightarrow b} |_{\tau_\mathrm{eff}},
\eeq
where we have neglected the second-order term $\mathcal{O}[\Gamma^2 \tau_1 \tau_2]$.
This shows that two short pulses are indeed equivalent to a single longer pulse with effective time duration given by the sum of the individual pulses,
$\tau_{\mathrm{eff}} = \tau_1 + \tau_2$,
as we could intuitively expect.

We now consider the the opposite limit, namely, two very long pulses in time (highly monochromatic), $\Delta_1 \ll \Gamma$ and $\Delta_2 \ll \Gamma$.
We have that
\beq
p^{(k)}_{a\rightarrow b} \approx r,
\eeq
which does not depend on $\Delta_k$.
Hence,
\beq
p^{(T,2)}_{a\rightarrow b} \approx (1-r) r + r,
\eeq
where the parameter $r$ is bounded to 
$r = 4\Gamma_a \Gamma_b /(\Gamma_a + \Gamma_b)^2 \leq 1$.

Depending on $r$, we find two possible scenarios.
If the two decay rates are identical ($\Gamma_a = \Gamma_b$, thus $r=1$), we have that 
$p^{(T,2)}_{a\rightarrow b} = r = p^{(T,1)}_{a\rightarrow b}$.
This shows that two pulses can be replaced by a single one again, as in the case of two short pulses.
However, in the far more typical case where $\Gamma_a \neq \Gamma_b$, we have that $r < 1$, so that
\beq
p^{(T,2)}_{a\rightarrow b} > r = p^{(T,1)}_{a\rightarrow b}.
\eeq
We see that a single long pulse saturates the transition probability to a value below unity ($p^{(1)}_{a\rightarrow b} = r < 1$), and the addition of a second pulse is the only way to improve driven self-organization.

For $0< r <1$, and considering $N > 2$ photons of arbitrary linewidths, we have that
\begin{align}
p^{(T,N)}_{a\rightarrow b} 
&=
\frac{r}{1+ \Delta_1/\Gamma}+
\sum_{k=2}^N \prod_{j=1}^{k-1} \left(1 - \frac{r}{1+ \Delta_j/\Gamma} \right) \frac{r}{1+ \Delta_k/\Gamma} \nn\\
&> \frac{r}{1+ \Delta_1/\Gamma},
\end{align}
which means that the degree of self-organization only increases with $N$.

Note that $p^{(T,N)}_{a\rightarrow b} \leq 1$ for $N\rightarrow \infty$, as we can see by taking the extreme limit where all the photons are highly monochromatic, thus maximally increasing the transition probability.
In that case, $\Delta_k / \Gamma \rightarrow 0$ for all $k$, so that
\begin{align}
p^{(T,N)}_{a\rightarrow b} 
&\approx r + r \sum_{k=1}^{N-1} (1-r)^k \nn \\
& = r + r \left( (1-r) \ \frac{1 - (1-r)^{N-1} }{r} \right) \nn \\
& = 1 - (1-r)^{N} \nn \\
& \leq 1.
\end{align}
We have used that, in the geometric series, 
$\sum_{n=1}^{M} a_1 q^{n-1} = a_1 (1-q^M) / (1-q)$, 
with $a_1 = q = 1-r$, and $M=N-1$.
So the series in our quantum model of $N$ cascaded photons converges, as it should.

\section{Conclusions}
We have investigated the validity of the dissipative adaptation in a quantum model where each $\Lambda$ atom in an ensemble is driven by a cascaded sequence of $N$ single-photon pulses.
We have found that, the more energy the atoms are given, the more organized they become, as reminiscent of the evolutionary dynamics of living systems.
Our model generalizes the previous ones where the quantum dissipative adaptation has been found, namely, its original proposal in Ref.\cite{qda}, as well as its applications in the self-replication of quantum artificial organisms \cite{qsr}, and the emergence of energy-seeking and energy-avoiding in classically-driven dissipative few-level quantum systems \cite{cp22}.
We have also found that, in the typical case where the atom has unequal decay rates, the level of organization achieved with two pulses matches that with a single one, provided the pair of photons are short enough; 
otherwise, the presence of multiple photons becomes a resource to raise the degree of self-organization.

As a perspective, we plan to study the case where the two pulses overlap.
This could lead to irreversible stimulated emissions, as shown in Ref.\cite{dvnjp}, eventually forcing the atom to go back to its initial state $\ket{a}$, thereby reducing the chances of self-organization to happen.
Understanding not only whether this is the case, but also how to compute work in that more complex scenario, as well as further testing the quantum dissipative adaptation hypothesis, are challenging and timely questions.

\begin{acknowledgements}
This work was supported by CNPq (Universal 402074/2023-8, and INCT-IQ 465469/2014-0), Brazil.
T.G. was supported by CAPES.
\end{acknowledgements}



%

%

%

%

%
\end{document}